\newcommand{\stm}{S\text{-}T_-}
\newcommand{\tm}{T_-}
\newcommand{\dst}{\Delta_{ST}}
\newcommand{\est}{\epsilon_{ST}}
\newcommand{\dstsq}{\langle\Delta_{ST}^{2}\rangle}
\newcommand{\ptime}{\tau_{w}} 
\newcommand{\PT}{\langle P_T \rangle}

\documentclass[%
reprint,amsmath,amssymb,aps,prl,superscriptaddress,longbibliography,
floatfix,
nobalancelastpage
]{revtex4-2}

\usepackage{graphicx}
\usepackage{dcolumn}
\usepackage{bm}
\usepackage{epstopdf}
\usepackage{float}
\usepackage{braket}
\usepackage{dsfont}
\usepackage{amsmath}
\usepackage{siunitx} 
\usepackage[italicdiff]{physics}
\usepackage[T1]{fontenc}
\usepackage[dvipsnames]{xcolor}
\usepackage{color,soul}
\usepackage{lipsum}
\usepackage[colorlinks=true, allcolors=blue]{hyperref}

\graphicspath{{figures/}}

\begin{document}

\title{The formation of a nuclear-spin dark state in silicon}

\author{Xinxin Cai}
\thanks{Present address: Department of Physics, Southern University of Science and Technology, Shenzhen, China}
\affiliation{Department of Physics and Astronomy, University of Rochester, Rochester, NY, 14627 USA}

\author{Habitamu Y. Walelign}
\affiliation{Department of Physics and Astronomy, University of Rochester, Rochester, NY, 14627 USA}


\author{John M. Nichol}
\email{john.nichol@rochester.edu}
\affiliation{Department of Physics and Astronomy, University of Rochester, Rochester, NY, 14627 USA}

\begin{abstract}
Silicon-based qubits are often made by trapping individual electrons in quantum dots defined by electric gates. Quantum information can then be stored using the spin states of the electrons. However, the nuclei of the surrounding atoms also have spin degrees of freedom that couple to the electron spin qubits and cause decoherence. The emergence of a nuclear-spin dark state has been predicted to reduce this coupling during dynamic nuclear polarization, when the electrons in the quantum dot drive the nuclei in the semiconductor into a decoupled state. Here, we report the formation of a nuclear-spin dark state in a gate-defined silicon double quantum dot. We show that, as expected, the transverse electron-nuclear coupling rapidly diminishes in the dark state, and that this state depends on the synchronized precession of the nuclear spins. Moreover, the dark state significantly reduces the relaxation rate between the singlet and triplet electronic spin states. This nuclear-spin dark state has potential applications as a quantum memory or in quantum sensing, and might enable increased polarization of nuclear spin ensembles.

\end{abstract}

\pacs{}

\maketitle


\section{Introduction}
Nuclear spins are abundant in condensed matter systems, and the ability to manipulate large ensembles of nuclear spins has led to many technological advances. Nuclear magnetic resonance, for example, has profoundly shaped science and medicine. In the quantum realm, nuclear spin ensembles often create decoherence for semiconductor qubits based on electron spins, and the ability to precisely control these nuclei would be beneficial for applications like quantum computing. Indeed, many ideas from nuclear magnetic resonance have been employed to mitigate this decoherence. One such approach, dynamic nuclear polarization (DNP), aims to suppress fluctuations in nuclear spin ensembles by fully polarizing the nuclei. Platforms including bulk solids~\cite{jacquinot1974polarization, maly2008dynamic}, gate-defined quantum dots~\cite{ono2004,Petta2008,Foletti2009}, and self-assembled quantum dots~\cite{chekhovich2017measurement,millington2024approaching} all offer routes to DNP, but achieving full polarization remains a challenge.
Significant theoretical research has focused on understanding why this is the case. Multiple lines of research have predicted that a collective nuclear-spin dark state, which has suppressed hyperfine coupling to the electrons, should emerge during DNP~
\cite{Imamoglu2003,Taylor2003,Brataas2011,brataas2012dynamical,Brataas2014,Gullans2010,Gullans2013,rudner2010phase}. Understanding the nuclear-spin dark state is crucial to improving DNP, and the dark state is also predicted to be an excellent quantum memory and resource for quantum sensing~\cite{Taylor2003,Taylor2003b,Zaporski2023}.

Despite the importance of the nuclear-spin dark state, only indirect evidence for its existence exists. Data from reconstructed spin states in self-assembled quantum dots are consistent with predicted dark-state properties~\cite{Gangloff2021}, and measurements of hole-spin polarization recovery in lead-halide perovskite materials have suggested the presence of the dark state~\cite{Kirstein2023}. A direct confirmation of the nuclear-spin dark state and tests of its properties would serve as important verifications for these predictions and point the way to harnessing this exotic state for various applications.

Here, we present evidence for the formation of the nuclear-spin dark state in a Si gate-defined quantum dot. We directly probe the hyperfine coupling between two electronic spin states (the singlet and polarized triplet),  through Landau-Zener sweeps, and we find that it sharply reduces during DNP, in agreement with theoretical predictions. We also observe that the dark state persists for about 1 ms, in agreement with nuclear spin inhomogeneous dephasing times in Si. We also verify that the dark state significantly reduces singlet-triplet relaxation. Together with these experiments, we construct numerical simulations that model the quantum-mechanical behavior of the electrons, as well as the semi-classical behavior of each of the roughly $10^4$ nuclear spins interacting with the electrons in our device. 

Our work verifies key properties of the nuclear-spin dark state, which has been predicted to appear in multiple systems, including gate-defined and self-assembled quantum dots, as well as solid-state materials. The ability to probe the dark state will improve our understanding of this collective state. In the future, this understanding could lead to increased levels of nuclear-spin polarization, which could in turn lead to increased electron-spin coherence times. Moreover, our work demonstrates that exotic nuclear spin states can occur in the very same devices that are used for quantum computing~\cite{Hanson2007,Zwanenburg2013,Burkard2023}, potentially opening the door to using such states as quantum memories or as resources for quantum sensing.

\section{Setup}

Most previous work on DNP in gate-defined quantum dots has focused on GaAs/AlGaAs devices~\cite{Petta2008,Foletti2009}. All stable isotopes of Ga and As have non-zero nuclear spin, enhancing the hyperfine interaction strength and efficiency of DNP. However, the high density of nuclear spins in these materials presents an obstacle to exploring collective spin states. The significantly lower density of nuclear spins in Si quantum dots (the most common spinful isotope, $^{29}$Si, occurs with about 5$\%$ natural abundance) means that fewer nuclei must change their orientation to realize the dark state and that nuclear spins possess longer coherence times. 
 
\begin{figure}[t]
\includegraphics{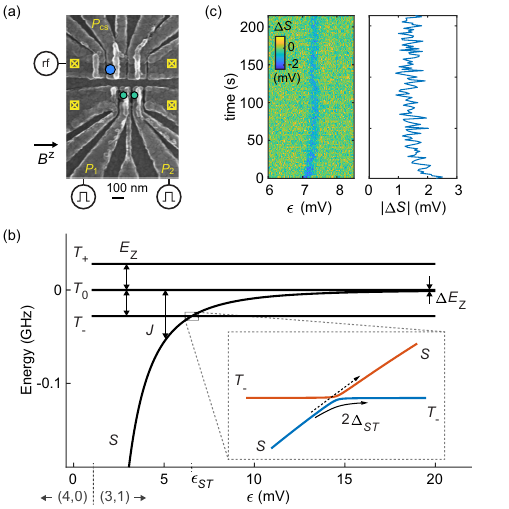}
	\caption{Experimental setup. 
        (a) Scanning electron micrograph of a device that is nominally identical to the one used. A double quantum dot is formed under plunger gates $P_1$ and $P_2$. A sensor dot is formed under gate $P_{cs}$. The magnetic field $B^z$ is applied parallel to the axis connecting the two dots, as indicated. 
        (b) Energy level diagram showing the relevant energy levels in the $(3,1)$ charge configuration. The variable $\epsilon$ is defined as a change in the plunger gate voltages such that $(\Delta V_{p1},\Delta V_{p2}) = (-\epsilon,\epsilon)$, with $\epsilon=0$ in the PSB region and $\epsilon=1$~mV near the $(4,0)$-$(3,1)$ charge transition. 
        Inset: an LZ sweep through $\est$, where the solid (dashed) arrow illustrates a passage through the avoided crossing that does (not) lead to a $S$ to $\tm$ transition.
        (c) Left panel: measurement of the $\stm$ avoided crossing position $\est$ over time. $\Delta S$ is the measured charge sensor signal relative to the mean signal value at $\epsilon<6.5$ mV. Negative values indicate non-zero triplet probability. Right panel: signal amplitude $|\Delta S|$ at $\est$. The data are extracted from the left panel, corresponding to the absolute value of the $\Delta S$ minima for each time value.
  }\label{fig:setup}
\end{figure}

In this work, we use a gate-defined double quantum dot fabricated on a natural Si/SiGe heterostructure [Fig.~\ref{fig:setup}(a)]. The electron occupancies of dot 1 and dot 2 are controlled by the positive voltages applied to the plunger gates, $V_{p1}$  and $V_{p2}$, respectively. We use the quantum dot beneath gate $P_{cs}$ as a sensor and measure its conductance via rf reflectometry~\cite{connors2020rapid} to read out the electronic state in the double dot. The device is cooled in a dilution refrigerator to a temperature of about 10~mK.

We operate the device near the $(4,0)$–$(3,1)$ charge transition, with 4(0) or 3(1) electrons in dot 1(2). In this configuration, two electrons in dot 1 always fill the lowest valley level as a singlet. The system dynamics can thus be described within a two-electron picture for the remaining ``valence'' electrons, which can form an effective singlet-triplet qubit~\cite{Petta2005,Harvey-Collard2017,connors2022charge}. We operate the device with four electrons, instead of the usual two for a singlet-triplet qubit, because using four electrons facilitates spin-blockade readout, even in the presence of low valley splittings~\cite{Harvey-Collard2017,harvey2018high,west2019gate,connors2020rapid}.

Figure~\ref{fig:setup}(b) shows the relevant energies of spin states in the $(3,1)$  charge configuration as a function of the detuning $\epsilon$. At a specific detuning $\est$, the exchange coupling $J$ equals the electronic Zeeman energy $E_Z=\bar{g} \mu_B B^z_{tot}$, where $\bar{g}\approx2$ is the electron $g$-factor in Si, and $B^z_{tot}$ is the total magnetic field experienced by the electrons, which includes both the externally applied field $B^z$ and the longitudinal hyperfine field. At $\est$, the spin singlet state, $|S\rangle=(\mid\uparrow\downarrow\rangle-\mid\downarrow\uparrow\rangle)/\sqrt{2}$, comes into resonance with the polarized triplet $|T_{-}\rangle = \mid\downarrow\downarrow\rangle$. The hyperfine interaction between the electron and nuclear spins couples these states. In particular, a difference in the transverse hyperfine field between the two dots couples $S$ and $\tm$ states, and the absolute value of this coupling is denoted by $\dst$~\cite{taylor2007relaxation}. A transverse hyperfine field couples $S$ and $T_-$, because these states differ in the $z$-component of their spin angular momentum. In general, a transverse magnetic field is thus required to transition from one state to the other. However, a uniform transverse field would flip the spins in both dots, leading to no coupling between these states. A difference in the transverse field is required to preferentially flip one spin, leading to a transition from $S$ to $T_-$. Deep in the (3,1) charge configuration, a difference in the longitudinal hyperfine field couples $S$ and $T_0$~\cite{taylor2007relaxation}.

All measurements presented in this paper begin with the state initialized as a singlet in $(4,0)$. Then, we manipulate the spin state of the valence electrons by pulsing $\epsilon$, typically near $\est$. To measure the device, we perform a Pauli spin blockade (PSB) readout via radiofrequency reflectometry~\cite{connors2020rapid} at $\epsilon = 0$ mV, allowing us to distinguish between the singlet and triplet configurations.

\section{Results}
We first demonstrate that DNP occurs in our device.  
In double quantum dots, DNP can be achieved by bringing $S$ and $\tm$ into resonance~\cite{Petta2008,Foletti2009}. A transition from $S$ to $\tm$ decreases the $z$-axis angular momentum of the electron spins by $\hbar$. In the absence of effects like spin-orbit coupling, angular momentum in the electron-nuclear system is conserved, and the $z$-axis angular momentum of the nuclear-spin system changes by $\hbar$. 

Figure~\ref{fig:setup}(c) shows the results of an experiment with the external magnetic field $B^z=1$~mT. We repeatedly initialize the electrons as a singlet and then pulse $\epsilon$ to different values near $\est$ for a fixed time of $4~\mu$s before the measurement. In the left panel, the thin line indicates the points at which the $\stm$ transition occurs \cite{Cai2023}.
We observe that $\est$ increases over time, corresponding to a reduction in $B^z_{tot}$. We have verified that this shift in $\est$ does not result from drifts in our superconducting magnet. This change in the total field is a hallmark of DNP in double quantum dots and has been observed previously in GaAs double quantum dots~\cite{Petta2008,Foletti2009}. As described in Methods, we also observe that the longitudinal magnetic Zeeman gradient $\Delta E_Z$ changes during a DNP process, consistent with results from GaAs devices.  We do not observe significant DNP effects at larger magnetic fields. This likely occurs because spin-orbit effects~\cite{Nichol2015,Cai2023}, which can quench DNP, become significant at higher fields.

\begin{figure}
\includegraphics{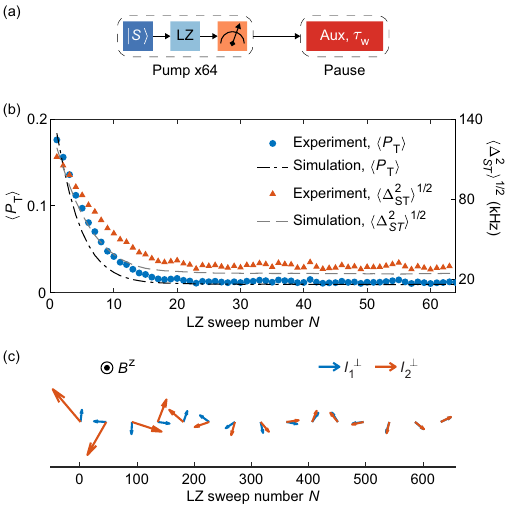}
	\caption{Driving into the dark state.
        (a) Pulse sequence for the experiment. Each pump cycle consists of a singlet load, an LZ sweep through $\est$, and a single-shot singlet-triplet measurement. A pause sequence is added after every 64 pump cycles, which pauses the pumping for $\ptime = 5.12$ ms.
        (b) Experimental data and numerical simulation of the LZ transition probability $\langle\it{P}_T\rangle$ (left axis) and the r.m.s.~$\stm$ coupling strength $\dstsq^{1/2}$ (right axis) as a function of LZ sweep number $N$.
        (c) Numerical results showing the synchronization of the nuclear spin precession around the external magnetic field $B^z$ between the two dots. The blue and red arrows denote the instantaneous orientation and relative amplitude of the in-plane components of $I_1$ and $I_2$ after the $N$th LZ sweep, where $\bm{I}_i$ is the average nuclear polarization in dot~$i$. For demonstration purposes, nuclear spin dephasing and relaxation are not included in the simulation of (c).
  }\label{fig:LZ}
\end{figure}

Curiously, the $\stm$ transition probability appears to decrease with time. This is revealed by the decreasing signal amplitude at $\est$ over time [Fig.~\ref{fig:setup}(c), right panel]. To explore this feature with better control, we perform DNP via Landau-Zener (LZ) sweeps through $\est$~\cite{Petta2008,Foletti2009}, [Fig.~\ref{fig:setup}(b)]. 
 We measure how the LZ transition probability depends on the number of sweeps. We repeat a ``pump'' cycle, consisting of a singlet initialization, an LZ sweep through $\est$, and a single-shot singlet-triplet measurement, 64 times [Fig.~\ref{fig:LZ}(a)]. During the LZ sweep, we sweep $\epsilon$ from 6.2 to 7.7 mV through $\est$ in $5$ $\mu$s, with a time interval $\tau_{w1}$ of $27~\mu$s to the next sweep. After this set of 64 pump cycles, we halt the pumping for $\tau_w = 5.12$ ms to allow the nuclear spin system to dephase and to implement auxiliary pulses to calibrate our measurement (see Methods). Figure~\ref{fig:LZ}(b) shows the triplet probability  $\langle\it{P}_T\rangle$ after each LZ sweep as a function of the sweep number, averaged over more than $10^4$ pump-pause cycles.

The experimental data reveal that $\langle P_{T}\rangle$ rapidly decreases with the sweep number and saturates at a small value [Fig.\ref{fig:LZ}(b)]. The remainder of this paper will explore this phenomenon. One possible explanation for this effect is that our singlet-triplet readout contrast diminishes with increased nuclear polarization. Such an effect has been observed in past work in GaAs quantum dots~\cite{Reilly2008,Barthel2012}. This effect occurs because the $T_0$ state relaxation rate increases significantly when the longitudinal magnetic gradient $\Delta E_Z$ becomes large, usually on the order of hundreds of MHz. However, this effect is unlikely to explain our data, because the typical $\Delta E_Z$ values are on the order of MHz in our device (see Extended Data Fig.~1). In addition, we measure the $T_-$ state, which does not experience increased relaxation at large $\Delta E_Z$. Another possible explanation for our data is that the process of DNP is so efficient that $\est$  moves outside of our LZ sweep window. As discussed further below, and as shown in Fig.~\ref{fig:elife}, this is also unlikely. 

A third explanation is the formation of the nuclear-spin dark state. The dark state should emerge during DNP as the nuclei are driven into a state with significantly reduced $\stm$ coupling~\cite{Brataas2011,Gullans2013}. In the case of a double quantum dot, the dark state is characterized by a vanishing net transverse hyperfine gradient ($2\dst$ in magnitude). Figure~\ref{fig:LZ}(b) shows the root-mean-square values of $\dst$ extracted from experimental data, using the analytical expression for the LZ transition probability \cite{Landau1932,Zener1932} as $\langle P_{T}(t)\rangle = \langle1 - \exp[-\frac{(2\pi)^2\dst^2(t)}{\beta}]\rangle\approx \frac{(2\pi)^2\langle\dst^2(t)\rangle}{\beta}$, where $\beta$ is the sweep rate [Table~\ref{tab:simpar}]. $\dstsq^{1/2}$ rapidly decreases with the LZ sweep number from 110 to 30 KHz.

The dark state can be understood in the following semiclassical picture. In thermal equilibrium, the average polarization along any direction of an ensemble of $N$ nuclear spins is approximately $\sqrt{N}$. Thus, the average imbalance in the transverse polarization between two dots with separate nuclear spin ensembles is also of order $\sqrt{N}$. During an LZ sweep, the quantum spin state of the electrons creates an effective magnetic field (the Knight field) that causes the difference of the transverse nuclear polarizations to rotate out of the $x$-$y$ plane and toward the $z$ direction. The reason for this behavior is the following. In brief, during an LZ sweep, the electronic spin state changes to the extent that a transverse difference field exists. If the electronic spin-state changes, the Knight field will acquire a transverse component. This transverse component will tend to rotate the nuclear polarizations from the $x$-$y$ plane to the $z$ axis. The transverse component of the Knight field has different signs for the two dots (see the Methods), so the difference in the transverse polarizations between the dots will be effectively rotated out of the plane. Thus, after about $\sqrt{N}$ LZ sweeps, the difference in the transverse polarization should vanish, leading to a scenario where $\dst=0$. In this ``dark state'', the transverse polarizations of the two dots are the same, and they undergo a synchronized precession around the external field.  

We perform numerical simulations of the full electron-nuclear system using the semiclassical techniques described in Refs.~\cite{Brataas2011,Brataas2014} and the parameters in Table~\ref{tab:simpar}. This approach combines a quantum description of the two electron spins with a semiclassical model of the coherent dynamics for each of the nuclear spins interacting with the electrons in the quantum dots. At each time during the LZ sweeps, we compute the quantum state of the electrons, the effective Knight field exerted by the electrons on the nuclear spins, and how the nuclei evolve in this Knight field. We account for nuclear spin dephasing and relaxation. A nuclear spin relaxation time of $T_{1,n}=5$~s is chosen to reproduce the saturation $\langle P_T\rangle$ value in Fig.~\ref{fig:LZ}(b), and is broadly consistent with other measurements we have made (see Extended Data Fig.~1). We also account for electrical noise experienced by the electrons. We consider 8000 and 4000 nuclei for dot~1 and dot~2, respectively, estimated according to the physical properties of the Si quantum dots. We use an hyperfine interaction strength of $\bar{A}=-0.052$ GHz, corresponding to a two-electron spin dephasing time of approximately 0.8~$\mu$s \cite{Assali2011}. The negative sign of $\bar{A}$ is determined by the signs of the electron and nuclear $g$-factors in Si. See Methods for further details on our simulations.

\begin{table}[t]
\caption{
\label{tab:simpar}
Parameters used for the simulations, where $N_{1(2)}$ is the number of nuclear spins in dot 1(2), $\bar{A}$ is the hyperfine interaction strength, $I$ is the amplitude of each nuclear spin, $\it{T}^*_{2,n}$ is the nuclear spin dephasing time, $\it{T}_{1,n}$ is the nuclear spin relaxation time, $\sigma_\epsilon$ describes the effect of electrical noise on $\epsilon$, and $\beta$ is the LZ sweep rate. We consider white noise for the nuclear spin dephasing unless otherwise indicated.} 
\begin{ruledtabular}
\begin{tabular}{ccccc}
$N_{1}$ & 8000                    & &     $\it{T}^*_{2,n}$ & 4.1 ms \\
$N_{2}$ & 4000                    & &     $\it{T}_{1,n}$ & 5 s\\
$\bar{A}$ & -0.052 GHz & &     $\sigma_\epsilon$ & $4.6\times10^{-4}$ GHz\\
$I$ & $1/2$            & &     $\beta$ & $2.8\times10^{-6}$ GHz/ns\\
\end{tabular}
\end{ruledtabular}
\end{table}

Figure~\ref{fig:LZ}(c) shows a simulated evolution of the nuclear spins during repeated LZ sweeps starting from a random state, confirming the semiclassical picture for the dark state formation. In particular, the difference in transverse polarization decreases in about $\sqrt{10^4}=10^2$ sweeps, consistent with our expectation that $\sqrt{N}$ sweeps are required to reach the dark state. Figure~\ref{fig:LZ}(b) shows simulations of our experiment, which agree quantitatively with our measurements. The residual $\dst$ is associated with the finite time interval $\tau_{w1}$ between LZ sweeps (see Methods). One surprising feature of these data, which is captured by the simulations, is that $\dst$ reduces over about 10 sweeps, while the expected number of sweeps to reach the dark state is about $10^2$, as shown in Fig.~\ref{fig:LZ}(c). The explanation for this effect is that over the roughly $10^4$ pump-pause cycles (each of which contains 64 LZ sweeps), a significant longitudinal nuclear polarization builds up. As the longitudinal polarization increases, the torque from the Knight field during LZ sweeps can more effectively rotate the transverse polarization to the $z$-direction (see Methods), and fewer sweeps are required to reach the dark state.

We next investigate what happens during the pause periods. The semiclassical picture of the dark state provided above suggests that the synchronized precession of the net polarizations in each dot underlies the vanishing electron-nuclear coupling. In this case, the primary mechanism for the nuclear spin system to exit the dark state involves inhomogeneous nuclear spin dephasing. One might intuitively expect that nuclear spin relaxation should spoil the dark state. While the nuclear spin $T_1$ time does limit the lifetime of the dark state, dephasing occurs more rapidly.  Within this picture, and starting from a random initial state, each dot has a net transverse nuclear polarization of order $\sqrt{N}$. Once the system enters the dark state after the LZ sweeps, the net transverse polarizations of the two dots have the same magnitude and phase. After the pause period, the net transverse polarizations of the dots have changed, because all of the $N$ nuclear spins within each dot have dephased through inhomogeneous broadening. This dephasing results in an effective new, random transverse polarization associated with each dot of order  $\sqrt{N}$. 

\begin{figure}
\includegraphics{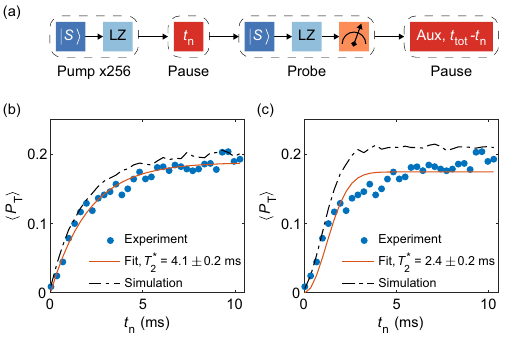}
	\caption{Dark state dephasing.
        (a) Pulse sequence for the measurement consists of 256 pump cycles, a pause for a variable time of $t_n$, a probe cycle to measure $P_T$, and an additional pause that maintains the total duration of the sequence. 
        (b) Measured transition probability $\langle P_T \rangle$ as a function of $t_n$ (filled circles). The data are fitted to the form $ P_0(1-\exp[-(t_n/T_c)])$ for nuclear spin dephasing due to white noise, with $P_0$ and $T_c$ as fitting parameters (solid line). The fitted value of $T_2^*=2T_c$ with the standard deviation is indicated. A numerical simulation considering white noise with the fitted $T_2^*$ time is also shown (dashed line).  
        (c) The same data are fitted to the form $P_0(1-\exp[-(t_n/T_c)^2])$ for dephasing due to quasi-static noise (solid line). The fitted value of $T_2^*=\sqrt{2}T_c$ with the standard deviation is indicated. The corresponding numerical simulation is shown (dashed line).  
  }\label{fig:dsdeph}
\end{figure}

We explore this effect using a ``pump-pause-probe-pause'' sequence, as shown in Fig.~\ref{fig:dsdeph}(a). The sequence starts with 256 pump cycles to drive the nuclear spins into the dark state. Then, we let the nuclei precess and dephase freely for a variable time of $t_n$ without pumping. Immediately afterward, we perform one more LZ sweep to measure $\langle P_T\rangle$. A second ``pause'' is applied at the end to compensate for the variation in $t_n$, to keep the total duration of the pulse sequence the same for the purpose of consistency. Each probability value is obtained from over $10^3$ repetitions of such a pulse sequence. As shown in Fig.~\ref{fig:dsdeph}b, the measured $\langle P_T\rangle$ increases with increasing $t_n$ (filled circles), revealing a recovery of the $\stm$ coupling over the timescale of a few milliseconds, consistent with previous measurements of dephasing times of $^{29}$Si nuclei in Si quantum-dot devices~\cite{pla2014coherent}.

To model these data, we consider the evolution of $\dst$ under the influence of nuclear spin dephasing. Assuming that at $t_n=0$, $\dst=0$, we analytically find that $\dstsq \propto 1-\exp[-(t_n/T_c)^\gamma]$. The characteristic time is closely related to the nuclear spin dephasing time, given by $T_c = T^*_{2,n}/\alpha$, with a factor $\alpha$ difference. Assuming that each nuclear spin experiences independent, Gaussian magnetic noise with a white power spectrum, $\dstsq$ follows an exponential function with $\gamma=1$ and $\alpha=2$. For quasi-static nuclear spin dephasing noise, $\dstsq$ follows a Gaussian function with $\gamma=2$ and $\alpha=\sqrt{2}$. By incorporating the form of $\dstsq$ into the expression for the LZ transition probability $\langle P_{T}(t)\rangle \approx \frac{(2\pi)^2\langle\dst^2(t)\rangle}{\beta}$, we fit the experimental data. In the case of white nuclear spin dephasing noise [Fig.~\ref{fig:dsdeph}(b)], the best-fit value of $T^*_{2,n}$ is 4.1 ms, while for quasistatic noise [Fig.~\ref{fig:dsdeph}(c)], the best-fit value is 2.4 ms.  In the figures, we also show numerical simulations with the fitted $T^*_{2,n}$ values. White dephasing noise, corresponding to exponential nuclear spin dephasing, appears to better describe the experimental data [Fig.~\ref{fig:dsdeph}(b)].

\begin{figure}
\includegraphics{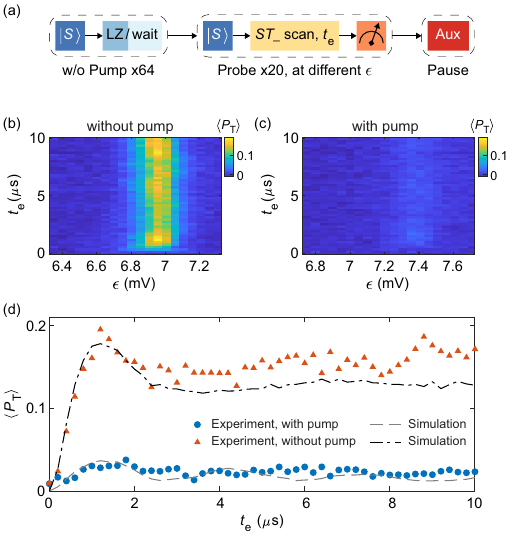}
	\caption{Decreasing singlet-triplet relaxation.
        (a) Pulse sequence for measurement with and without pumping. In the case without pumping, the LZ sweep in the pump cycles is replaced by a wait pulse at $\epsilon=0$. For the probe cycles, we initialize the singlet state and pulse to different values of $\epsilon$ for a variable evolution time $t_e$.
        (b) Measured $\langle P_T \rangle$ for $\epsilon$ values around $\est$ in the absence of pumping.
        (c) Measured $\langle P_T \rangle$ for $\epsilon$ values around $\est$ with pumping.
        (d) $\langle P_T \rangle$ at $\est$ as a function of $t_e$. The experimental data (filled triangles and circles) are extracted from (b) and (c), respectively, corresponding to the $\langle P_T \rangle$ maximum for each $t_e$ value. Numerical simulations for both cases are presented. 
  }\label{fig:elife}
\end{figure}

One major potential benefit of the nuclear-spin dark state is the possibility of reduced relaxation between the singlet and triplet electronic states, due to the reduced hyperfine coupling. We discuss the possibility of reduced electron-spin dephasing in the Discussion section. In the singlet-triplet basis we use in this work, a relaxation reduction is expected because the nuclear spin operators couple to the electron raising and lowering operators, as opposed to the $\sigma^z$ operator (see Methods for more details on the full Hamiltonian). 

To test this idea,  we measure the relaxation of the singlet state near $\est$ with and without DNP. 
In the former case, the pulse sequence starts with 64 pump cycles to form the dark state, followed by a ``probe'' cycle, as shown in Fig.~\ref{fig:elife}(a). In the case without pumping, we use an identical pulse sequence but replace each LZ sweep in the pump cycle by a wait at $\epsilon=0$~mV, away from $\est$. In each probe cycle, with the electronic state initialized as a singlet, we pulse $\epsilon$ abruptly to a value near $\est$, wait at that $\epsilon$ for a variable time $t_e$, allowing the state to relax, and then abruptly pulse $\epsilon$ back to 0 mV for the measurement. Again, each probability value is obtained from over $10^3$ repeated measurements. 

Figures~\ref{fig:elife}(b) and (c) show the measured $\langle P_T \rangle$ after the wait near $\est$ for variable time $t_e$ without and with pumping, respectively. 
Without pumping, $\langle P_T \rangle$ at $\est$ increases over a time-scale of about a microsecond and saturates between 0.1 and 0.2, as indicated by the yellow region in Fig.~\ref{fig:elife}(b) and filled triangles in Fig.~\ref{fig:elife}(d). The relatively low saturation value reflects both the relatively large voltage resolution used in this measurement, as well as the presence of detuning noise in our system. Both effects can result in a situation where the experimental detuning is not exactly $\est$, leading to a reduced transition probability. 

With pumping, however, $\langle P_T\rangle$ remains nearly zero up to at least 10 $\mu$s [Fig.~\ref{fig:elife}(d)], indicating strongly suppressed singlet-triplet relaxation. In principle, $\langle P_T\rangle$ should remain small for up to the lifetime of the dark state, and future work will focus on confirming this hypothesis. As above, our numerical simulations agree well with our data [Fig.~\ref{fig:elife}(d)]. We also note that even in the dark state, Fig.~\ref{fig:elife}(c) shows that $\est$ still falls within our LZ sweep window, as discussed above.


\section{Discussion}
We have presented evidence for a nuclear-spin dark state in silicon, including the rapid reduction in the transverse hyperfine singlet-triplet coupling strength with repeated Landau-Zener sweeps. Our measurements of the dark-state lifetime are consistent with nuclear-spin dephasing, and we find significantly reduced electron-spin relaxation in the dark state. Together with these measurements, we also developed semiclassical numerical simulations that compute the behavior of all electron and nuclear spins involved in our device.  

More broadly, our work has confirmed important predictions about the nuclear-spin dark state, which has been predicted to occur in multiple experimental systems, like gate-defined quantum dots, self-assembled quantum dots, and solid-state materials. One area of future research will focus on using our improved understanding of the dark state to enable more efficient DNP in a variety of systems. For example, further experimental work is required to determine the nuclear spin relaxation time $T_{1,n}$. Experimentally, we have observed that this relaxation time seems to depend on the exact configuration of the nuclear spins, as well as the pulses used to manipulate the electrons. Based on our simulations, it appears that the level of polarization achieved in the nuclear spin system depends sensitively on the relaxation rate. Efficient DNP will also depend on the development of ways to ``brighten'' the dark state to enable continued polarization~\cite{Imamoglu2003}. Finally, spin-orbit coupling is also predicted to control the presence of a topological phase transition related to the formation of the dark state and DNP~\cite{rudner2010phase}.

With more efficient DNP, it may be possible to improve the coherence of electron-spin qubits.   Here, we have demonstrated a reduction in the singlet-triplet relaxation rate in the dark state. However, coherence times of various spin qubits are also expected to improve if full polarization can be achieved, or in regimes where LZ sweeps can act as weak measurements~\cite{ribeiro2009nuclear}. In either case, the effects of the dark state are critical to understand in order to realize improved coherence associated with these procedures. Even in highly-optimized spin systems like isotopically-purified silicon spin qubits, residual hyperfine coupling limits coherence times, and increased polarization could benefit such qubits.  

Another critical area of future research will seek to understand quantum effects in the nuclear spin system, and how these effects impact the electron spins. In addition, we plan to explore how to use dark states as a quantum memory, resource for sensing, or other quantum applications~\cite{Taylor2003b,Zaporski2023}. 

During the completion of this work, we became aware of related work exploring nuclear-spin dark states as quantum memories~\cite{appel2024many}.

\section{Acknowledgments}
We thank Lisa F. Edge of HRL Laboratories, LLC. for the epitaxial growth of the SiGe material and 
Elliot J. Connors for device fabrication. This work was sponsored by the Army Research Office under grants W911NF-17-1-0260 (J.M.N.) and W911NF-19-1-0167 (J.M.N.), and by the National Science Foundation under grant OMA 1936250 (X.C. and J.M.N.) and DMR-1941673 (H.Y.W. and J.M.N.). The views and conclusions contained in this document are those of the authors and should not be interpreted as representing the official policies, either expressed or implied, of the Army Research Office or the U.S. Government. The U.S. Government is authorized to reproduce and distribute reprints for Government purposes notwithstanding any copyright notation herein.

\section{Author Contributions}
X.C., H.Y.W., and J.M.N. conceptualized the experiment and wrote the manuscript. X.C. and H.Y.W. conducted the investigation.  X.C. and J.M.N. developed the simulation code.
J.M.N. supervised the project. 

\section{Competing Interests}
The authors declare no competing interests. 

\section{Methods}

\subsection{DNP in Silicon}

In this section, we demonstrate the measurement of $\Delta E_Z$, the difference in the longitudinal magnetic Zeeman energy between the two dots, with and without pumping. We set $B^z=1~$mT and apply the pulse sequences in Extended Data Fig.~\ref{fig:dbz}(a). In the case with pumping, the pulse sequence starts with 16 pump cycles. During each LZ sweep, we sweep $\epsilon$ from 6.2 to 7.7~mV in 5~$\mu$s, with a time interval of 20~$\mu$s to the next sweep. In the case without pumping, each LZ sweep is replaced by a wait at $\epsilon=0$~mV, away from $\est$.

We measure $\Delta E_Z$ through coherent oscillations between the $S$ and $T_0$ states at a large $\epsilon$. The oscillation frequency corresponds to the energy splitting between the eigenstates with suppressed exchange interaction, $f = \sqrt{J_r^2+(\Delta E_Z)^2}$, where $J_r$ is the residual exchange. The top panel in Extended Data Fig.~\ref{fig:dbz}(b) shows the experimental result, where each vertical line is a repetition of the oscillation measurement. The pumping starts at the beginning, and there is a clear development of coherent oscillations over repetitions. After the pumping stops at around 158~s, those oscillations disappear over time. Bottom panel in Extended Data Fig.~\ref{fig:dbz}(b) displays the result of the fast Fourier transform (FFT) of the data. The FFT peaks reveal that the oscillation frequency increases to above 2~MHz with pumping and drops back to around 0.8~MHz after the pumping stops. The residual frequency is probably due to the nonvanishing $J_r$. Based on previous measurements of this device~\cite{Cai2023}, we expect residual exchange values of about 1~MHz or less. The change in oscillation frequency reflects the change in $\Delta E_Z$. This change in $\Delta E_Z$ is another hallmark of DNP in double quantum dots \cite{Foletti2009}.

\subsection{$\ptime$ and $\tau_{w1}$ dependence}

Extended Data Figure~\ref{fig:tau2}(a) is a control experiment corresponding to Fig.~2(b), showing the dependence on the pause time $\ptime$. We use a ``pump-pause'' sequence similar to Fig.~2(a) and repeat the sequence over $10^4$ times to obtain the probability value. The data from this experiment reveal that the $\PT$ values at the beginning of the pump cycles significantly decrease with reducing $\ptime$. For short $\ptime$, the measured $\PT$ is small for all $N$ (blue filled circles in Fig.~\ref{fig:tau2}(a). 

The data in Extended Data Fig.~\ref{fig:tau2}(a) is consistent with the picture of the dark state formation. Without any pause, the nuclear spins will remain in the dark state once formed. Averaging over the roughly $10^4$ repetitions, we anticipate obtaining the saturation $\PT$ value for all $N$. Once a pause period is added, during which the dark state dephases, a recovery of $\PT$ value is expected. At the start of this pause period, the net transverse nuclear polarizations of each dot are approximately equal in the dark state. During the pause period, the net transverse polarizations change randomly as a result of inhomogeneous dephasing of the individual nuclear spins, leading to an effective new, random transverse polarization associated with each dot and letting the nuclear spin system exit the dark state.

Extended Data Figure~\ref{fig:tau2}(b) shows a control experiment where the time interval between subsequent LZ sweeps, $\tau_{w1}$, is varied. The saturation $\PT$  value reflects the difference in transverse hyperfine fields between the dots that reappears during $\tau_{w1}$, as a result of nuclear-spin dephasing. Longer $\tau_{w1}$ will lead to more dephasing and thus a larger saturation value. Data in Extended Data Fig.~\ref{fig:tau2}(b) align with the expectation.

\subsection{Pulse sequence for the pauses}

In this section, we describe the pulse sequence employed to generate pauses of pumping in the experiments discussed in the main text. 

During the pause periods, we continue sending pulses but replace each LZ sweep by a wait at $\epsilon=0$~mV, away from $\est$, to halt pumping. We also insert auxiliary pulses to calibrate our measurement, denoted as ``Aux'' in the pulse-sequence diagrams. An auxiliary pulse involves initializing a random spin state, by first pulsing to (3,0) to empty dot 2 and then to (3,1) to load either a singlet or triplet \cite{connors2020rapid}, before a single-shot singlet-triplet measurement. For the experiments in the main text, we keep all pulses to be $32~\mu$s in length for consistency. To be specific, for the measurement in Fig.~2(b), the pause period consists of 64 auxiliary pulses and 96 wait-at-zero pulses, with a total duration of 5.12~ms; for the measurement in Fig.~3, the first pause period consists a variable number of wait-at-zero pulses (between 0-320) to pause pumping for different $t_n$, and the second pause period consists of 64 auxiliary pulses and a complementary number of wait-at-zero pulses (between 320-0) to maintain the total duration of the pause periods ($t_{tot} = 12.288$~ms). For the measurements in Fig.~4, since no specific pause period is required, the pause is made of only 64 auxiliary pulses with a total duration of 2.048~ms. Similarly, for the measurement in Extended Data Fig.~\ref{fig:dbz}, the pause period consists of only 64 auxiliary pulses, with a minor difference in the settings that the pulse length is $25~\mu$s.

\subsection{Simulation}

Numerical simulations in this work involve the integration of the Schr$\ddot{\text{o}}$dinger equation with a time-dependent Hamiltonian $H^{ST}$ for electron spin dynamics, alongside integrating the precession equations around an effective magnetic field $\bm{\Delta_i}$ to model the nuclear spin dynamics. 

In the $\{\ket{\tm},\ket{S}\}$ subspace, the Hamiltonian of the electron spin state is given by
\begin{equation}
 	H^{ST} = \left( \begin{array}{c c}
 	\Delta  E_{ST} + v^z	&	v^+ \\
 	v^-	&	0	
 	\end{array} \right).
\label{eq:H}
 \end{equation}
Here, $\Delta E_{ST} = J(\epsilon) - \bar{g}\mu_B B^z$ is the energy difference between $\tm$ and $S$ in the absence of longitudinal nuclear polarization. We experimentally determine the voltage dependence of the exchange interaction $J(\epsilon)$  through a standard technique named spin funnel measurements \cite{Petta2005,Cai2023}. For the Zeeman-energy term, we use $\bar{g}=2$ and $B^z = 1$~mT. During an LZ sweep, $\Delta E_{ST}$ changes approximately from 0.004 to -0.01~GHz over a duration of 5~$\mu$s. In simulations, we approximate that $\Delta E_{ST}$ changes linearly with $\epsilon$ within the LZ sweep window, corresponding to a constant sweep rate of $\beta=d(\Delta E_{ST})/dt = 2.8\times10^{-6}~$GHz/ns.

$v^\pm$ and $v^z$ are nuclear-spin dependent terms. $v^\pm$ is the difference in transverse hyperfine fluctuations between the two dots, which gives rise to the $\stm$ avoided crossing with $\dst=|v^\pm|$, in the form of 
 \begin{equation}
     \begin{split}
          v^{\pm} = \frac{\bar{A}}{2}(I_{1}^{\pm}-I_{2}^{\pm}), 
     \end{split}
     \label{eq:vin}
 \end{equation}
 where $\bar{A}$ is the electron-nuclear interaction strength. $\bm{I}_{i} = \frac{1}{N_i}\sum_j\bm{I}_{ij}$ is the average angular momentum of the nuclei in dot $i$, where $j$ denotes individual nucleus site in the dot. $I_{i}^{\pm}= (I_{i}^{x}\pm i I_{i}^{y})/\sqrt{2}$ are the transverse components. The
 $v^z$ term corresponds to the total longitudinal hyperfine field in the two dots, given by
 \begin{equation}
     \begin{split}
          v^{z} = -\frac{\bar{A}}{2}(I_{1}^{z}+I_{2}^{z}). 
     \end{split}
     \label{eq:vz}
 \end{equation} 
In the simulations of this work, the longitudinal nuclear polarization achieved is between 10$\%$ and 20$\%$. 
 
The dynamics of each of the nuclear spins are described by 
 \begin{equation}
     \begin{split}
           \hbar \frac{d\bm{I}_{ij}}{dt} = \bm{\Delta}_{i}\times\bm{I}_{ij}, 
     \end{split}
\label{eq:nsd}
 \end{equation}
 where the effective magnetic field $\bm{\Delta}_i$ acting on the $j$th nucleus in dot $i$ depends on the electron dynamics described by the singlet $c_S$ and triplet $c_{\tm}$ amplitudes of the electron wave function, 
 \begin{equation}
\begin{split}
\Delta_{i}^+ & = \bar{A} c_{S}^*c_{\tm} (-1)^{i-1}/(2N_i), \\
\Delta_{i}^- & = \bar{A} c_{\tm}^*c_{S} (-1)^{i-1}/(2N_i), \\
\Delta_{i}^z & = -\bar{A} |c_{\tm}|^2/(2N_i)  - g_n\mu_N B^z,
\end{split}
\label{eq:Knight1}
\end{equation}
where $i=1,2$. The term $g_n\mu_N B^z$ is the nuclear procession frequency around the external magnetic field. For $^{29}$Si, the geomagnetic ratio is $g_n\mu_N = -8.46$ MHz/T. 
  
At each time during the LZ sweeps, we compute $c_S$ and $c_{\tm}$ from the Schr$\ddot{\text{o}}$dinger equation,
 \begin{equation}
     \begin{split}
          H^{ST} \binom{c_{\tm}}{c_S} = i\hbar \partial_t \binom{c_{\tm}}{c_S}, 
     \end{split}
 \end{equation}
to simulate the evolution of electron spins, and we compute $\bm{\Delta}_i$ and $\bm{I}_{ik}$ using Eqs.~\ref{eq:Knight1} and \ref{eq:nsd} to simulate the evolution of nuclear spins. We use an integration time step of 10~ns. 

We consider 8000 and 4000 nuclei for dot 1 and 2, respectively. Based on the size of the (3,1) charge stability diagram and the lever arms of the plunger gates and assuming the self-capacitance of a flat disk, we roughly estimate the size of the two dots to be 46 and 35~nm in radius. Assuming a disk thickness of $4$~nm, we estimate the number of $^{29}$Si atoms in each dot based on the lattice constant of Si and the abundance of $^{29}$Si.

We set $\bar{A}=-0.052$ GHz. In simulations, when all nuclear spins are randomly distributed, this magnitude gives rise to a two-electron spin dephasing time of about 0.8~$\mu$s, which is broadly consistent with our other measurements. The negative prefactor is needed to match the direction of the $\est$ change observed in the experiment, and it aligns with the signs of $g$-factors in Si. 

For numerical simulations in Fig.~2, each of nuclear spins is initialized with a random orientation and an amplitude of $I=1/2$. The simulation results in Figs.~3 and 4 are averaged over 32 runs with different initial random states. 

The simulations include the nuclear spin relaxation and dephasing. We simulate the dephasing by assuming that each nuclear spin experiences an independent magnetic noise $\delta B^z$ on top of the external magnetic field $B^z$, either a white noise or a quasi-static noise as discussed in the main text. We simulate the nuclear spin relaxation by randomly re-initializing nuclear spins with a characteristic time of $T_{1,n}$. We also incorporate electrical noise into our simulations by generating quasi-static, Gaussian-distributed noise for $\Delta E_{ST}$ with a r.m.s.~value of $\sigma_\epsilon = 4.6\times10^{-4}~$GHz, corresponding to a voltage fluctuation of 49~$\mu$V.

\subsection{Theoretical model}
\label{sec:TM}

In this section, we derive the forms of $v^{\pm}$, $v^{z}$, and $\bm{\Delta}_i$ from the theoretical model provided by Brataas and Rashba \cite{Brataas2011,Brataas2014}. 

The Hamiltonian of the hyperfine electron-nuclear interaction is given by
\begin{equation}
H_{hf} = \bar{A}\sum_k \sum_{l=1}^2 \delta(\bm{R}_k - \bm{r}_l)(\bm{I}_k\cdot \bm{S}_l),
\label{eq:Hhf}
\end{equation}
where $l$ and $k$ represent the electrons and nuclear spins, respectively. 

In the $\{|\tm\rangle,|S\rangle\}$ subspace, the two-electron wave functions of the basis states are 
\begin{equation}
\begin{split}
\Psi_S(1,2) & = \psi_S(\bm{r}_1,\bm{r}_2)\chi_S(\bm{s}_1,\bm{s}_2), \\
\Psi_{\tm}(1,2) & = \psi_{\tm}(\bm{r}_1,\bm{r}_2)\chi_{\tm}(\bm{s}_1,\bm{s}_2),
\end{split}
\label{eq:wavefun}
\end{equation}
where the orbital wavefunction $\psi_S$ ($\psi_{\tm}$) and the spin-wavefunction $\chi_{\tm}$ ($\chi_{S}$) are symmetric (antisymmetric) under particle exchange. 

Eqs.~\ref{eq:Hhf} and \ref{eq:wavefun} yield the following matrix elements:
\begin{equation}
\begin{split}
v^z & = \langle \tm | H_{hf} | \tm\rangle  = -\bar{A} \sum_k \zeta_k {I}_k^z , \\
v^+ & = \langle \tm | H_{hf} | S\rangle  = \bar{A} \sum_k \rho_k \frac{ I_k^x + iI_k^y }{\sqrt{2}}, \\
\end{split} 
\label{eq:v}
\end{equation}
where 
\begin{equation}
\begin{split}
\zeta_k & = \int_{r} \psi_{\tm}^*(\bm{r},\bm{R}_k)\psi_{\tm} (\bm{r},\bm{R}_k) \mathrm{d}r, \\
\rho_k & = - \int_{r} \psi_{\tm}^*(\bm{r},\bm{R}_k)\psi_{S} (\bm{r},\bm{R}_k) \mathrm{d}r. \\
\end{split} \label{eq:coeff1}
\end{equation}
 
For nuclear spin dynamics, the effective driving fields $\bm{\Delta}_i$ in Eq.~\ref{eq:nsd} depend on the singlet $c_S$ and triplet $c_{\tm}$ amplitudes of the electron wave function, $\Psi = c_{\tm} \Psi_{\tm} + c_S \Psi_S$. By computing $\langle \Psi | H_{hf} | \Psi \rangle$, we obtain
\begin{equation}
\begin{split}
\Delta_{k}^+ & = \frac{(\Delta_{k}^x +i\Delta_{k}^y)}{\sqrt{2}} = \bar{A}\rho_{k}   c_{S}^*c_{\tm}, \\
\Delta_{k}^- & = \frac{(\Delta_{k}^x -i\Delta_{k}^y)}{\sqrt{2}} 
= \bar{A}\rho_{k}  c_{\tm}^*c_{S}, \\
\Delta_{k}^z & = -\bar{A}\zeta_{k} |c_{\tm}|^2  - g_n\mu_N B^z,
\end{split}
\label{eq:Knight}
\end{equation}
 where $\Delta_k^z$ has also included the effect of the external magnetic field $B^z$, and $g_n\mu_N$ is the geomagnetic ratio of the nuclei. 

In this work, we consider non-overlapping box-like electronic wavefunctions, meaning that each electron interacts homogeneously with the nuclear spins in its own dot. Hence, Eq.~\ref{eq:coeff1} becomes
\begin{equation}
\begin{split}
\zeta_k=\frac{1}{2N_1}~\text{and}~\rho_k=\frac{1}{2N_1},~\text{for dot~1} \\
\zeta_k=\frac{1}{2N_2}~\text{and}~\rho_k=-\frac{1}{2N_2},~\text{for dot~2} 
\end{split}\label{eq:coeff2}
\end{equation}
Combining with Eq.~\ref{eq:v} and Eq.~\ref{eq:Knight}, we obtain the forms of $v^{\pm}$, $v^{z}$, and $\bm{\Delta}_i$  in Eqs.~\ref{eq:vin}, \ref{eq:vz} and \ref{eq:Knight1}, respectively.

\subsection{Number of LZ sweeps required for the dark state}
The data of Fig.~2(b) in the main text show that the number of sweeps required to enter the dark state is about 10, instead of the expected value of about 10$^2$. Here we explain why. The simulation of Fig.~2(b) in the main text are averaged over 8192 pump-pause cycles, each with 64 LZ sweeps. During this process, a significant (10-20$\%$) longitudinal polarization builds up. To understand the effect of this polarization, let $I_1^x$, $I_1^y$, and $I_1^z$ represent the components of the average nuclear angular momentum in dot 1, and let $I_1=\sqrt{(I_1^x)^2+(I_1^y)^2+(I_1^z)^2}$ and $I_1^{\perp}=\sqrt{(I_1)^2-(I_1^z)^2}$. Assuming uniform hyperfine couplings and neglecting nuclear-spin relaxation and dephasing during a single LZ sweep, $I_1$ is conserved, so $dI_1^{\perp}/dI_1^z=-I_1^z/I_1^{\perp}$. If the electronic state changes from singlet to triplet with probability $P_{LZ}$, the $z$-component of the average nuclear angular momentum of dot 1 will change proportionally to $P_{LZ}$, since angular momentum in the combined electron-nuclear system is also conserved.  Thus, during a single LZ sweep, $\Delta I_1^{\perp} = -(I_1^z/I_1^{\perp}) \Delta I_1^z \propto -(I_1^z/I_1^{\perp}) P_{LZ}$. Thus, LZ sweeps are more effective at reducing the transverse polarization when the initial state is polarized. The same arguments apply also to dot 2. 

\subsection{Electron-spin lifetime simulation}

Extended Data Figure~\ref{fig:elifetime_sim} show the numerical simulations corresponding to Figs.~4(b-c) in the main text. The simulations presented in Fig.~4(d) are extracted from Extended Data Figs.~\ref{fig:elifetime_sim}(a) and (b), respectively, corresponding to the $\PT$ maximum for each $t_e$ value. In Extended Data Fig.~\ref{fig:elifetime_sim}, we use an $\epsilon$ step of $0.05$~mV, the same as in the experiment. The relatively large voltage resolution acts as an effective detuning noise in the experiment, with a magnitude about the same as $\sigma_\epsilon$. We cannot fully distinguish the effects of these two.

A feature that is captured by both simulations and experiments is an increase in $\est$. The $\stm$ transitions occur at larger $\epsilon$ with pumping in comparison to the case without it. This increase corresponds to a reduction in $B^z_{tot}$ due to the longitudinal nuclear polarization, as has been discussed in the main text. We emphasize that even with pumping, $\est$ remains within our LZ sweep window of 6.2-7.7~mV. Specifically, $\est\approx7.2~$mV in the simulation [Extended Data Fig.~\ref{fig:elifetime_sim}(b)], and $\est\approx7.4~$mV in the experiment [Fig.~4(c)]. We speculate that the small deviation in these $\est$ values has to do with the exact form of $\Delta E_{ST}(\epsilon)$ and because $d(\Delta E_{ST})/d\epsilon=dJ/d\epsilon$ is not truly a constant within the sweep window, as we assume in simulations, but slightly decreases with increasing $\epsilon$.

%

\section{Data Availability}
The processed data that support the plots are available at https://doi.org/10.5281/zenodo.12171443. The raw data are available from the corresponding author upon request.

\section{Code availability}
The code used to generate numerical simulations in this work is available at https://doi.org/10.5281/zenodo.12171443.

\renewcommand{\theequation}{S\arabic{equation}}
\renewcommand{\figurename}{Extended Data Fig.}
\setcounter{figure}{0}    

\onecolumngrid
\section{Extended Data}

\begin{figure}[H]
\centering
\includegraphics{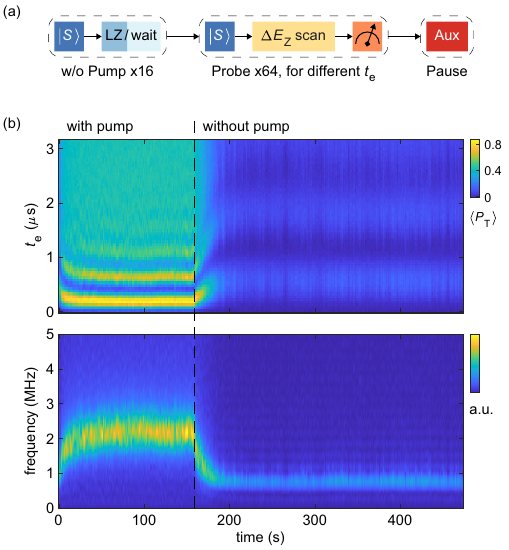}
	\caption{Measurement of longitudinal hyperfine gradient. 
        (a) Pulse sequence for the measurement with and without pumping. During the probe cycles, we initialize the singlet state, pulse to a fixed $\epsilon$ value deep in $(3,1)$ for different evolution times $t_e$, and measure $P_T$. 
        (b) Top panel: repeated measurement of $S\text{-}T_0$ oscillations at $\epsilon=30$~mV as a function of elapsed time. Bottom panel: absolute value of the fast Fourier transform of the data, showing the change in the oscillation frequency. Pumping starts at the beginning and stops after around 158~s, as indicated by the vertical dashed line.
  }\label{fig:dbz}
\end{figure}

\begin{figure}[H]
\centering
\includegraphics{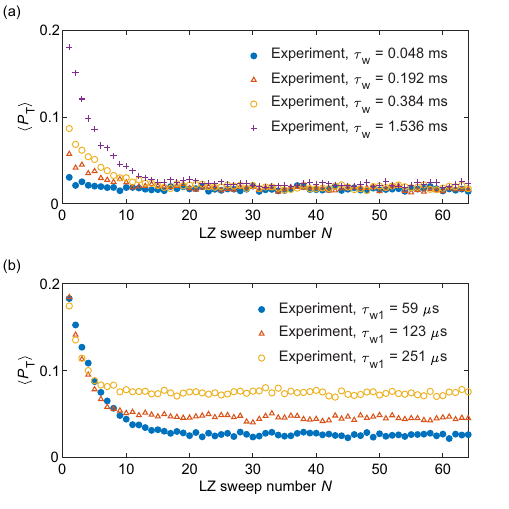}
	\caption{Control experiments corresponding to
Fig.~2(b) in the main text. (a) Measured $\PT$ as a function of LZ sweep number $N$, with different      values of $\ptime$ as indicated. We use the pulse sequence in Fig.~2(a), with minor adjustments in the parameters: during each LZ sweep, $\epsilon$ is swept from 6.1 to 7.6 mV over $5~\mu$s, with a time interval of $43~\mu$s to the next sweep. (b) Measured $\PT$ as a function of $N$, with different values of $\tau_{w1}$ as indicated. All other parameters remain the same as those used in the experiment of Fig.~2(b).
  }\label{fig:tau2}
\end{figure}

\begin{figure}[H]
\centering
\includegraphics{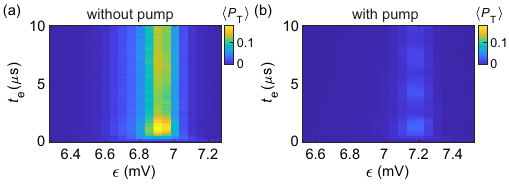}%
	\caption{Numerical simulations corresponding to Figs.~4(b-c) in the main text. 
  }\label{fig:elifetime_sim}
\end{figure}


\end{document}